# A BIAS CORRECTION FOR THE MINIMUM ERROR RATE IN CROSS-VALIDATION

By Ryan J. Tibshirani[1] and Robert Tibshirani[2]

*Stanford University and Stanford University*

Tuning parameters in supervised learning problems are often estimated by cross-validation. The minimum value of the cross-validation error can be biased downward as an estimate of the test error at that same value of the tuning parameter. We propose a simple method for the estimation of this bias that uses information from the cross-validation process. As a result, it requires essentially no additional computation. We apply our bias estimate to a number of popular classifiers in various settings, and examine its performance.

**1. Introduction.** Cross-validation is widely used in regression and classification problems to choose the value of a "tuning parameter" in a prediction model. By training and testing the model on separate subsets of the data, we get an idea of the model's prediction strength as a function of the tuning parameter, and we choose the parameter value to minimize the CV error curve. This estimate admits many nice properties [see Stone (1977) for a discussion of asymptotic consistency and efficiency] and works well in practice.

However, the minimum CV error itself tends to be too optimistic as an estimate of true prediction error. Many have noticed this downward bias in the minimum error rate. Breiman et al. (1984) acknowledge this bias in the context of classification and regression trees. Efron (2008) discusses this problem in the setting $p \gg n$, and employs an empirical Bayes method, which does not involve cross-validation in the choice of tuning parameters, to avoid such a bias. However, the proposed algorithm requires an initial

Received November 2008; revised November 2008.
[1]Supported by a National Science Foundation Vertical Integration of Graduate Research and Education fellowship.
[2]Supported in part by National Science Foundation Grant DMS-99-71405 and National Institutes of Health Contract N01-HV-28183.
*Key words and phrases.* Cross-validation, prediction error estimation, optimism estimation.

choice for a "target error rate," which complicates matters by introducing another tuning parameter. Varma and Simon (2006) suggest a method using "nested" cross-validation to estimate the true error rate. This essentially amounts to doing a cross-validation procedure for every data point, and is hence impractical in settings where cross-validation is computationally expensive.

We propose a bias correction for the minimum CV error rate in $K$-fold cross-validation. It is computed directly from the individual error curves from each fold and, hence, does not require a significant amount of additional computation.

Figure 1 shows an example. The data come from the laboratory of Dr. Pat Brown of Stanford, consisting of gene expression measurements over 4718 genes on 128 patient samples, 88 from healthy tissues and 40 from CNS tumors. We randomly divided the data in half, into training and test samples, and applied the nearest shrunken centroids classifier Tibshirani et al. (2001) with 10-fold cross-validation, using the pamr package in the R language. The figure shows the CV curve, with its minimum at 23 genes, achieving a CV error rate of 4.7%. The test error at 23 genes is 8%. The estimate of the CV bias, using the method described in this paper, is 2.7%, yielding an adjusted error of $4.7 + 2.7 = 7.4\%$. Over 100 repeats of this experiment, the average test error was 7.8%, and the average adjusted CV error was 7.3%.

In this paper we study the CV bias problem and examine the accuracy of our proposed adjustment on simulated data. These examples suggest that the bias is larger when the signal-to-noise ratio is lower, a fact also noted by Efron (2008). We also provide a short theoretical section examining the expectation of the bias when there is no signal at all.

**2. Model selection using cross-validation.** Suppose we observe $n$ independent and identically distributed points $(x_i, y_i)$, where $x_i = (x_{i1}, \ldots, x_{ip})$ is a vector of predictors, and $y_i$ is a response (this can be real-valued or discrete). From this "training" data we estimate a prediction model $\hat{f}(x)$ for $y$, and we have a loss function $L(y, \hat{f}(x))$ that measures the error between $y$ and $\hat{f}(x)$. Typically, this is

$$L(y, \hat{f}(x)) = (y - \hat{f}(x))^2 \qquad \text{squared error}$$

for regression, and

$$L(y, \hat{f}(x)) = 1\{y \neq \hat{f}(x)\} \qquad \text{0–1 loss}$$

for classification.

An important quantity is the expected prediction error $\mathrm{E}[L(y_0, \hat{f}(x_0))]$ (also called expected test error). This is the expected value of the loss when predicting an independent data point $(x_0, y_0)$, drawn from the same distribution as our training data. The expectation is over all that is random [namely, the model $\hat{f}$ and the test point $(x_0, y_0)$].

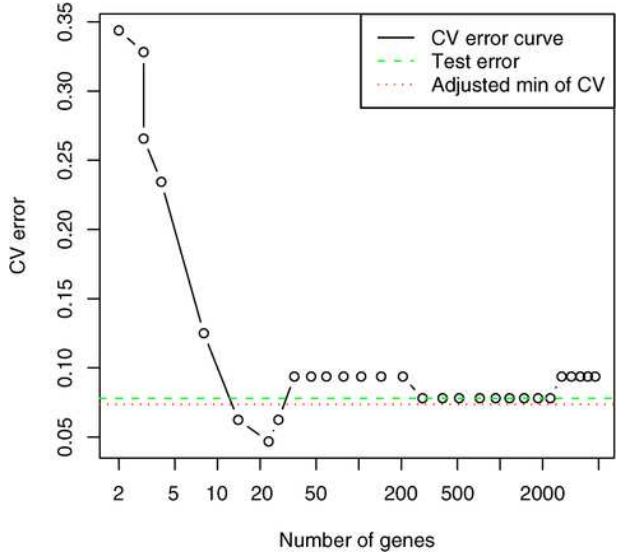

Fig. 1. *Brown microarray cancer data: the CV error curve is minimized at 23 genes, achieving a CV error of* 0.047. *Meanwhile, the test error at 23 genes is* 0.08, *drawn as a dashed line. The proposed bias estimate is* 0.027, *giving an adjusted error of* $0.047 + 0.027 = 0.074$, *drawn as a dotted line.*

Suppose that our prediction model depends on a parameter $\theta$, that is, $\hat{f}(x) = \hat{f}(x, \theta)$. We want to select $\theta$ based on the training set $(x_i, y_i), i = 1, \ldots, n$, in order to minimize the expected prediction error.

One of the simplest and most popular methods for doing this is $K$-fold cross-validation. We first split our data $(x_i, y_i)$ into $K$ equal parts. Then for each $k = 1, \ldots, K$, we remove the $k$th part from our data set and fit a model $\hat{f}^{-k}(x, \theta)$. Let $C_k$ be the indices of observations in the $k$th fold. The cross-validation estimate of the expected test error is

$$\text{(1)} \qquad \text{CV}(\theta) = \frac{1}{n} \sum_{k=1}^{K} \sum_{i \in C_k} L(y_i, \hat{f}^{-k}(x_i, \theta)).$$

Recall that $\hat{f}^{-k}(x_i, \theta)$ is a function of $\theta$, so we compute $\text{CV}(\theta)$ over a grid of parameter values $\theta_1, \ldots, \theta_t$, and choose the minimizer $\hat{\theta}$ to be our parameter estimate. We call $\text{CV}(\theta)$ the "CV error curve."

**3. Bias correction.** We would like to estimate the expected test error using $\hat{f}(x,\hat{\theta})$, namely,

$$\mathrm{Err} = \mathrm{E}[L(y_0, \hat{f}(x_0, \hat{\theta}))].$$

The naive estimate is $\mathrm{CV}(\hat{\theta})$, having bias

(2) $$\mathrm{Bias} = \mathrm{Err} - \mathrm{CV}(\hat{\theta}).$$

This is likely to be positive, since $\hat{\theta}$ was chosen because it minimizes $\mathrm{CV}(\theta)$.

Let $n_k$ be the number of observations in the $k$th fold, and define

$$e_k(\theta) = \frac{1}{n_k} \sum_{i \in C_k} L(y_i, \hat{f}^{-k}(x_i, \theta)).$$

This is the error curve computed from the predictions in the $k$th fold.

Our estimate uses the difference between the value of $e_k$ at $\hat{\theta}$ and its minimum to mimic the bias in cross-validation. Specifically, we propose the following estimate:

(3) $$\widehat{\mathrm{Bias}} = \frac{1}{K} \sum_{k=1}^{K} [e_k(\hat{\theta}) - e_k(\hat{\theta}_k)],$$

where $\hat{\theta}_k$ is the minimizer of $e_k(\theta)$. Note that this estimate uses only quantities that have already been computed for the CV estimate (1), and requires no new model fitting. Since $\widehat{\mathrm{Bias}}$ is a mean over $K$ folds, we can also use the standard error of the mean as an approximate estimate for its standard deviation.

The adjusted estimate of test error is $\mathrm{CV}(\hat{\theta}) + \widehat{\mathrm{Bias}}$. Note that if the fold sizes are equal, then $\mathrm{CV}(\hat{\theta}) = \frac{1}{K} \sum_{k=1}^{K} e_k(\hat{\theta})$ and the adjusted estimate of test error is

$$\mathrm{CV}(\hat{\theta}) + \widehat{\mathrm{Bias}} = 2\,\mathrm{CV}(\hat{\theta}) - \frac{1}{K} \sum_{k=1}^{K} e_k(\hat{\theta}_k).$$

The intuitive motivation for the estimate $\widehat{\mathrm{Bias}}$ is as follows: first, $e_k(\hat{\theta}_k) \approx \mathrm{CV}(\hat{\theta})$ since both are error curves evaluated at their minima; the latter uses all $K$ folds, while the former uses just fold $k$. Second, for fixed $\theta$, cross-validation error estimates the expected test error, so that $e_k(\theta) \approx \mathrm{E}[L(y, \hat{f}(x, \theta))]$. Thus, $e_k(\hat{\theta}) \approx \mathrm{Err}$.

The second analogy is not perfect: $\mathrm{Err} = \mathrm{E}[L(y, \hat{f}(x, \hat{\theta}))]$, where $(x,y)$ is stochastically independent of the training data, and hence of $\hat{\theta}$. In contrast, the terms in $e_k(\hat{\theta})$ are $L(y_i, \hat{f}^{-k}(x_i, \hat{\theta})), i \in C_k$; here $(x_i, y_i)$ has some dependence on $\hat{\theta}$ since $\hat{\theta}$ is chosen to minimize the validation error across all folds, including the $k$th one. To remove this dependence, one would have to carry out a new cross-validation for each of the $K$ original folds, which is much more computationally expensive.

There is a similarity between the bias estimate in (3) and bootstrap estimates of bias in Efron (1979) and Efron and Tibshirani (1993). Suppose that we have data $z = (z_1, z_2, \ldots, z_n)$ and a statistic $s(z)$. Let $z^{*1}, z^{*2}, \ldots, z^{*B}$ be bootstrap samples each of size $n$ drawn with replacement from $z$. Then the bootstrap estimate of bias is

$$\widehat{\text{Bias}}_{\text{boot}} = \frac{1}{B} \sum_{b=1}^{B} [s(z^{*b}) - s(z)]. \tag{4}$$

Suppose that $s(z)$ is a functional statistic and hence can be written as $t(\hat{F})$, where $\hat{F}$ is the empirical distribution function. Then $\widehat{\text{Bias}}_{\text{boot}}$ approximates $E_F[t(\hat{F})] - t(F)$, the expected bias in the original statistic as an estimate of the true parameter $t(F)$.

Now to estimate the quantity Bias in (2), we could apply the bootstrap estimate in (4). This would entail drawing bootstrap samples and computing a new cross-validation curve from each sample. Then we would compute the difference between the minimum of the curve and the value of curve at the training set minimizer. In detail, let $\text{CV}(z^*, \hat{\theta}(\tilde{z}))$ be the value of the cross-validation curve computed on the dataset $z^*$ and evaluated at $\hat{\theta}(\tilde{z})$, the minimizer for the CV curve computed on dataset $\tilde{z}$. Then the bootstrap estimate of bias can be expressed as

$$\frac{1}{B} \sum_{b=1}^{B} [\text{CV}(z^{*b}, \hat{\theta}(z)) - \text{CV}(z^{*b}, \hat{\theta}(z^{*b}))]. \tag{5}$$

The computation of this estimate is expensive, requiring $B$ $K$-fold cross-validations, where $B$ is typically 100 or more. The estimate in $\widehat{\text{Bias}}$ in (3) finesses this by using the original cross-validation folds to approximate the bias in place of the bootstrap samples.

In the next section we examine the performance of our estimate in various contexts.

**4. Application to simulated data.** We carried out a simulation study to examine the size of the CV Bias, and the accuracy of our proposed adjustment (3). The data were generated as standard Gaussian in two settings: $p < n$ ($n = 400, p = 100$) and $p \gg n$ ($n = 40, p = 1000$). There were two classes of equal size. For each of these we created two settings: "no signal," in which the class labels were independent of the features, and "signal," where the mean of the first 10% of the features was shifted to be 0.5 units higher in class 2.

In each of these settings we applied five different classifiers: LDA (linear discriminant analysis), SVM (linear support vector machines), CART (classification and regression trees), KNN ($K$-nearest neighbors), and GBM (gradient boosting machines). In the $p \gg n$ setting, the LDA solution is not of full rank, so we used diagonal linear discriminant analysis with soft-thresholding of the centroids, known as nearest shrunken centroids (NSC). Table 1 shows the mean of the test error, minimum CV error (using 10-fold CV), true bias, and estimated bias, over 100 simulations. The standard errors are given in brackets.

TABLE 1
*Results for proposed bias correction for the minimum CV error, using 10-fold cross-validation. Shown are mean and standard error over 100 simulations, for five different classifiers*

|  | Method | Min CV error | Test error | Adjusted CV error |
|---|---|---|---|---|
|  |  | $p < n$ |  |  |
| No signal | LDA | 0.503 (0.003) | 0.5 | 0.503 (0.003) |
|  | SVM | 0.485 (0.003) | 0.5 | 0.511 (0.004) |
|  | CART | 0.474 (0.003) | 0.5 | 0.510 (0.004) |
|  | KNN | 0.473 (0.002) | 0.5 | 0.524 (0.003) |
|  | GBM | 0.475 (0.003) | 0.5 | 0.520 (0.003) |
| Signal | LDA | 0.290 (0.003) | 0.284 (0.001) | 0.290 (0.003) |
|  | SVM | 0.257 (0.003) | 0.260 (0.001) | 0.279 (0.003) |
|  | CART | 0.356 (0.003) | 0.378 (0.002) | 0.384 (0.003) |
|  | KNN | 0.291 (0.003) | 0.284 (0.002) | 0.305 (0.004) |
|  | GBM | 0.269 (0.002) | 0.272 (0.002) | 0.288 (0.003) |
|  |  | $p \gg n$ |  |  |
| No signal | NSC | 0.384 (0.009) | 0.5 | 0.511 (0.012) |
|  | SVM | 0.475 (0.009) | 0.5 | 0.498 (0.010) |
|  | CART | 0.498 (0.011) | 0.5 | 0.500 (0.011) |
|  | KNN | 0.430 (0.007) | 0.5 | 0.577 (0.009) |
|  | GBM | 0.432 (0.010) | 0.5 | 0.552 (0.012) |
| Signal | NSC | 0.106 (0.006) | 0.136 (0.004) | 0.152 (0.008) |
|  | SVM | 0.142 (0.007) | 0.138 (0.003) | 0.157 (0.008) |
|  | CART | 0.432 (0.012) | 0.432 (0.004) | 0.437 (0.012) |
|  | KNN | 0.200 (0.007) | 0.251 (0.005) | 0.297 (0.010) |
|  | GBM | 0.233 (0.008) | 0.276 (0.006) | 0.307 (0.010) |

We see that the bias tends to larger in the "no signal" case, and varies significantly depending on the classifier. And it seems to be sizable only when $p \gg n$. The bias adjustment is quite accurate in most cases, except for the KNN and GBM classifiers when $p \gg N$, when it is too large. With only 40 observations, 10-fold CV has just four observations in each fold, and this may cause erratic behavior for these highly nonlinear classifiers. Table 2 shows the results for KNN and GBM when $p \gg N$, with 5-fold CV. Here the bias estimate is more accurate, but is still slightly too large.

**5. Nonnegativity of the bias.** Recall Section 3, where we introduced Bias = Err − CV($\hat{\theta}$), and our estimate $\widehat{\text{Bias}}$. It follows from the definition that $\widehat{\text{Bias}} \geq 0$ always. We show that for classification problems, E[Bias] $\geq 0$ when there is no signal.

THEOREM 1. *Suppose that there is no true signal, so that $y_0$ is stochastically independent of $x_0$. Suppose also that we are in the classification setting, and $y_0 = 1, \ldots, G$ with equal probability. Finally suppose that the loss is 0–1, $L(y, \hat{f}(x)) = 1\{y \neq \hat{f}(x)\}$. Then $E[CV(\hat{\theta})] \leq \text{Err}$.*

PROOF. The proof is quite straightforward. Well Err $= 1 - P(y_0 = \hat{f}(x_0, \hat{\theta}))$, where $\hat{f}(\cdot, \hat{\theta})$ is fit on the training examples $(x_1, y_1), \ldots, (x_n, y_n)$. Suppose that marginally $P(\hat{f}(x_0, \hat{\theta}) = j) = p_j$, for $j = 1, \ldots, G$. Then, by independence,

$$P(y_0 = \hat{f}(x_0, \hat{\theta})) = \sum_j P(y_0 = \hat{f}(x_0, \hat{\theta}) = j) = \sum_j \frac{1}{G} p_j = \frac{1}{G},$$

so Err $= \frac{G-1}{G}$. By the same argument, $E[CV(\theta)] = \frac{G-1}{G}$ for any fixed $\theta$. Therefore,

$$E[CV(\hat{\theta})] = E\left[\min_i CV(\theta_i)\right] \leq E[CV(\theta_1)] = \frac{G-1}{G},$$

which completes the proof. □

Now suppose that there is no signal and we are in the regression setting with squared error loss, $L(y, \hat{f}(x)) = (y - \hat{f}(x))^2$. We conjecture that indeed $E[CV(\hat{\theta})] \leq \text{Err}$ for a fairly general class of models $\hat{f}$.

Let $(\tilde{x}_1, \tilde{y}_1), \ldots, (\tilde{x}_n, \tilde{y}_n)$ denote $n$ test points, independent of the training data and drawn from the same distribution. Consider doing cross-validation on the test set in order to determine a value for $\theta$ (just treating the test data like it were training data). That is, define

$$\widetilde{CV}(\theta) = \frac{1}{n} \sum_{k=1}^{K} \sum_{i \in C_k} (\tilde{y}_i - \tilde{f}^{-k}(\tilde{x}_i, \theta))^2,$$

TABLE 2
*Results for KNN and GBM when $p \gg N$, with 5-fold cross-validation*

| Classifier | Setting | Min CV error | Test error | Adjusted CV error |
|---|---|---|---|---|
| KNN | No signal | 0.430 (0.007) | 0.5 | 0.524 (0.009) |
| KNN | Signal | 0.213 (0.007) | 0.253 (0.005) | 0.281 (0.009) |
| GBM | No signal | 0.425 (0.008) | 0.5 | 0.511 (0.010) |
| GBM | Signal | 0.265 (0.008) | 0.289 (0.007) | 0.325 (0.010) |

where $\tilde{f}^{-k}$ is fit on all test examples $(\tilde{x}_i, \tilde{y}_i)$ except those in the $k$th fold. Let $\tilde{\theta}$ be the minimizer of $\widetilde{\text{CV}}(\theta)$ over $\theta_1, \ldots, \theta_t$. Then

$$\text{E}[\text{CV}(\hat{\theta})] = \text{E}[\widetilde{\text{CV}}(\tilde{\theta})] \leq \text{E}[\widetilde{\text{CV}}(\hat{\theta})],$$

where the first step is true by symmetry, and the second is true by definition of $\tilde{\theta}$. But (assuming for notational simplicity that $1 \in C_1$) $\text{E}[\widetilde{\text{CV}}(\hat{\theta})] = \text{E}[(\tilde{y}_1 - \tilde{f}^{-1}(\tilde{x}_1, \hat{\theta}))^2]$, and we conjecture that

(6) $$\text{E}[(\tilde{y}_1 - \tilde{f}^{-1}(\tilde{x}_1, \hat{\theta}))^2] = \text{E}[(\tilde{y}_1 - \hat{f}^{-1}(\tilde{x}_1, \hat{\theta}))^2].$$

Intuitively, since there is no signal, $\tilde{f}(\cdot, \hat{\theta})$ and $\hat{f}(\cdot, \hat{\theta})$ should predict equally well against a new example $(\tilde{x}_1, \tilde{y}_1)$, because $\hat{\theta}$ should not have any real relation to predictive strength.

For example, if we are doing ridge regression with $p = 1$ and $K = n$ (leave-one-out CV), and we assume that each $x_i = \tilde{x}_i$ is fixed (nonrandom), then we can write out the model $\hat{f}^{-k}(\cdot, \theta)$ explicitly. In this case, we can show (6) is equivalent to showing

$$\text{E}[y_1|\hat{\theta}] = \text{E}[y_1], \qquad \text{E}[y_1^2|\hat{\theta}] = \text{E}[y_1^2] \quad \text{and} \quad \text{E}[y_1 y_2|\hat{\theta}] = \text{E}[y_1]\text{E}[y_2].$$

In words, the mean and variance of $y_1$ are unchanged by conditioning on $\hat{\theta}$, and $y_1, y_2$ are conditionally independent given $\hat{\theta}$. These certainly seem true when looking at simulations, but are hard to prove rigorously because of the complicated relationship between the $y_i$ and $\hat{\theta}$.

Similarly, we conjecture that

(7) $$\text{E}[(\tilde{y}_1 - \hat{f}^{-1}(\tilde{x}_1, \hat{\theta}))^2] = \text{E}[(\tilde{y}_1 - \hat{f}(\tilde{x}_1, \hat{\theta}))^2],$$

because there is no signal. If we could show (6) and (7), then we would have $\text{E}[\text{CV}(\hat{\theta})] \leq \text{E}[(\tilde{y}_1 - \hat{f}(\tilde{x}_1, \hat{\theta}))^2] = \text{Err}$.

**6. Discussion.** We have proposed a simple estimate of the bias of the minimum error rate in cross-validation. It is easy to compute, requiring essentially no additional computation after the initial cross-validation. Our studies indicate that it is reasonably accurate in general. We also found that the bias itself is only an issue when $p \gg N$ and its magnitude varies considerably depending on the classifier. For this reason, it can be misleading to compare the CV error rates when choosing between models (e.g., choosing between NSC and SVM); in this situation the bias estimate is very important.

**Acknowledgments.** The authors would like to thank the Editor for helpful comments that led to improvements in the paper.

Department of Statistics
Stanford University
Stanford, California 94305-5405
E-mail: ryantibs@stanford.edu

HRP Redwood Bldg
Stanford University
Stanford, California 94305-5405
E-mail: tibs@stanford.edu